# Team Power Dynamics and Team Impact:
# New Perspectives on Scientific Collaboration using Career Age as a Proxy for Team Power

Huimin Xu

*School of Information, University of Texas at Austin, Austin, TX 78701, USA*

Yi Bu

*Department of Information Management, Peking University, Beijing 100871, China*

Meijun Liu

*Institute for Global Public Policy, Fudan University, Shanghai 200433, China*

Chenwei Zhang

*Faculty of Education, The University of Hong Kong, Hong Kong 999077, China*

Mengyi Sun

*Department of Ecology and Evolutionary Biology,* University of Michigan*, MI 48109, USA*

Yi Zhang

*Faculty of Engineering and IT, University of Technology Sydney, NSW 2007, Australia*

Eric Meyer

*School of Information, University of Texas at Austin, Austin, TX 78701, USA*

Eduardo Salas

*Department of Psychological Science, Rice University, Houston, TX 77005, USA*

Ying Ding

*School of Information, University of Texas at Austin, Austin, TX 78701, USA*

**Abstract**: Power dynamics influence every aspect of scientific collaboration. Team power dynamics can be measured by team power level and team power hierarchy. Team power level is conceptualized as the average level of the possession of resources, expertise, or decision-making authorities of a team. Team power hierarchy represents the vertical differences of the possessions of resources in a team. In Science of Science, few studies have looked at scientific collaboration from the perspective of team power dynamics. This research examines how team power dynamics affect team impact to fill the research gap. In this research, all co-authors of one publication are treated as one team. Team power level and team power hierarchy of one team are measured by the mean and Gini index of career age of co-authors in this team. Team impact is quantified by citations of a paper authored by this team. By analyzing over 7.7 million teams from Science (e.g., Computer Science, Physics), Social Sciences (e.g., Sociology, Library & Information Science), and Arts & Humanities (e.g., Art), we find that flat team structure is associated with higher team impact, especially when teams have high team power level. These findings have been repeated in all five disciplines except Art, and are consistent in various types of teams from Computer Science including teams from industry or academia, teams with different gender groups, teams with geographical contrast, and teams with distinct size.

**Keywords**: Team Power Dynamics; Team Power Level, Team Power Hierarchy; Flat Structure; Team Impact; Scientific Collaboration

# INTRODUCTION

Any understanding of teamwork is imperfect if the notion of power is ignored. Power is manifested through cooperation (French & Raven, 1959, Blau, 1964, 1977) and influences every step of collaboration. As collaboration becomes common, power dynamics are consistently presented in team activities (Dahl, 1957). Practicing modern science requires an unprecedented level of teamwork (Bozeman & Youtie, 2017). Science of Science researchers study teamwork by examining coauthorship (He et al., 2013), diversity (Bu et al., 2018), team size (Larivière et al., 2014; Wu et al., 2019), gender (Larivière, et al., 2013), and international collaboration (Wagner &

Leydesdorff, 2005), but few pay attention to the role of team power dynamics in scientific teams. In this research, we examine how team power dynamics affect team impact to fill the research gap.

Team power can be defined as whether one team member has the capacity to influence others as well as the capacity to be uninfluenced by others in the same team (Magee & Galinsky, 2008). In team collaboration, team power dynamics can be measured by the combination of team power level and team power hierarchy (Greer et al., 2017). Team power level is conceptualized as the average control of resources, average expertise level, or average position of a team (Greer et al., 2011; Greer & Van Kleef, 2010; Groysberg et al., 2011). Team power level can be high (e.g., manager teams) or low (e.g., entry-level employee teams). Team power hierarchy is defined as an implicit or explicit rank order of individuals or groups with respect to a valued social dimension (Magee & Galinsky, 2008). It represents vertical differences between team members in their possessions of team resources and their levels of authority for decision making (Greer et al., 2018). Team power hierarchy can be vertical (e.g., top down or concentrated power) or flat (e.g., everyone holds similar levels of power). Team power hierarchy can help or hurt team effectiveness in different contexts and tasks (Anderson & Brown, 2010; Bunderson et al., 2016). It can improve cooperation, coordination, and role clarity, but it can also breed inequality and resentment.

Age has been historically emphasized as the basis of power (Linton, 1936). When people of different ages work together, older people would normally be given higher power or status in the hierarchy (Harrison & Klein, 2007; Gingras et al. 2008). Age-based power determines the productivity, impact, and division of labor in teams. As scientists become more senior, their positions in the hierarchy tend to rise, they gain access to more resources, and thus can increase their productivity and impact (Merton, 1973). Younger scholars often make more specific "technical" contributions while senior ones focus on the big picture and "conceptual" tasks (Larivière et al., 2016). The academic tenure system is consistent with career age that when faculty member's career age progresses, the corresponding academic rank and authority tend to increase (Craft et al., 2016). While there are individual exceptions, generally career age can be treated as a reasonable proxy to represent power in scientific teams. Other proxies for power could be job positions in corporations (Greer & Van Kleef, 2010), pay and participation in sports competition (Halevy et al., 2012, Bloom, 1999), and speaking turns in group discussion (Woolley et al., 2010).

In this paper, we use the mean and Gini coefficient of all team members' career ages as measures for team power level and team power hierarchy. A scientific team with high team power level consists of a majority of senior scientists with longer career ages. By contrast, teams with more newcomers and junior researchers are those with low team power level.

This paper starts a new way to understand scientific collaboration by studying the relationships between team power dynamics and team impact. It detects patterns in these relationships and verifies them in several selected fields including Computer Science and Physics from Science, Sociology and Library & Information Science from Social Sciences, and Art from Arts & Humanities.  It uses career age as a proxy for power and takes the related factors about gender, context (in terms of teams from academia or industry), country, productivity, team size, and time into consideration. In this paper, all co-authors of one publication is treated as one team. Team power level and team power hierarchy of one team are measured by using the mean and Gini coefficients of career age of all co-authors in this team. Team impact is quantified by citations of a paper authored by this team. Through the analyses of over 7.7 million teams from five different fields, we find general and consistent patterns that teams with low team power hierarchy are cited more than teams with high team power hierarchy when team power level increases. Flat structure is correlated with higher team impact.

## RELATED WORK

***Team Power Dynamics: Team Power Level and Team Power Hierarchy***
Team power dynamics have been extensively studied in different contexts including organizations, sports, and military. Greer and Van Kleef (2010) explored the interactive effect between team power level and team power hierarchy on team success in commercial companies. They found that team power hierarchy is useful for teams with low power level, but problematic for teams with high power level. In low-power teams, individuals are more likely to accept hierarchy because clear roles could help them know what to do (Goodwin et al., 2018). However, in high-power teams, members prefer balanced power without hierarchy because members with high power are more prone to be competitive and sensitive when power differences exist within the team. Greer et al. (2011) conducted field studies within a financial company to test the effect of power level on team success. Their results indicate that a team made up of high-power people (e.g., a management team) performs worse than a team consisting of low power people (e.g., secretarial teams) when collaborating in a logic puzzle game. In sports, team power hierarchy has distinct influence in different ball games. Halevy et al. (2012) measured team power hierarchy using the differences in salaries and playing time in matches for professional NBA basketball teams. By controlling the team power level (e.g., the average salaries of teams), they concluded team power hierarchy can increase the winning percentage by enhancing intragroup coordination. Conversely in baseball teams, team power hierarchy embodied by pay distribution has a negative influence on on-field performance (Bloom, 1999). To illustrate the differences in results, Ronay et al. (2012) suggested that the extent of task interdependence might influence the function of team power hierarchy. Compared with relatively independent baseball games, basketball games rely more on coordination and cooperation, and thus team members might regard team power hierarchy as legitimate and acceptable. In the military, for lower power level teams, like aircrews or rifle squads, a highly hierarchical setting with military leaders is necessary (Goodwin et al., 2018). In science, team power has not been extensively addressed (Satterly et al., 2018).

*Team Power Dynamics and Related Factors*

Extensive evidence suggests that gender disparity leads to difference in productivity, collaboration, and impact in Science of Science (Larivière et al., 2013). By analyzing the positions of authorship in Science and Humanities, West et al. (2013) found that when men and women collaborate together, men tend to take the leading role as the first or last author. The specific division of labor for different genders has been identified by Robinson-Garcia et al. (2020) that women perform specialized tasks (e.g., experiments and analyzing data) by contrast men perform supporting and leading tasks (e.g., designing and writing). Meanwhile, females have a homophily preference to work among the same genders and countries (Uhly et al., 2017) which can lead to diminishment in creative ideas and fewer citations (Larivière et al., 2013). In addition, shortened career lengths and high dropout rates are associated with the underrepresentation of women in productivity and impact (Huang et al., 2020). Research suggests that the proportion of different genders influences how power is shared within teams, and determines the ultimate team success in different tasks, such as puzzle, brainstorming, negotiating, and class projects (Bear & Woolley, 2011; Berdahl & Anderson, 2005; Woolley et al., 2008).

Universities and companies have distinct characteristics in terms of working environment and approaches to knowledge disclosure (Chai & Shih, 2016). Scientists have the freedom to choose research directions and share knowledge through peer-reviewed papers. By contrast, firms are driven by economic benefits therefore researchers must apply patents to protect their intellectual properties. The collaboration between academia and industry has advantages and disadvantages (Bikard et al., 2019; Rybnicek & Konigsgruber, 2019), which include differing views about open science (Czarnitzki et al., 2015) and demands of commercialization (Toole & Czarnitzki, 2010). Tijssen (2004) found the steady increase in industry and academia collaboration and Lebeau et al. (2008) discovered university-industry articles receive more citations than pure university or industry papers in Canadian institutions.

The growth of international collaborations is a steady trend (Gazni et al., 2012). When working in the culturally diverse teams, collaborators can exchange different skills and knowledge (Abbasi & Jaafari, 2013). Literature shows that cross-cultural collaboration is more positively related to high citations (Hsiehchen et al., 2015; Wagner et al., 2017). But, working with people from different countries faces challenges, such as divergent culture, language, communication, and thinking styles (Brett et al., 2006). Wagner et al. (2019) found that papers' novelty decreases but the conventionality increases when the number of countries increases. In summary, while gender, industry and academia collaboration, and culture can significantly impact team collaboration and team success, the current literature does not investigate the wholistic view of these factors and how their effects can affect team power dynamics on team impact. This paper provides quantitative measures of team power and systematic evaluations of these factors.

*Team Collaboration in Science of Science*
Science of Science researchers have produced rich literature about team collaboration from perspectives of demographic, geographic, and network characteristics. Demographic variables include the composition of gender, age, ethnicity, and tenure in teams. The overarching finding is that the diversity of team composition has a positive effect on team performance. For example, researchers found that the proportion of females is positively related to equal conversational turn-taking during the cooperation process (Woolley et al., 2008). The diversity of career age within teams can bring 0.59 more citations each year (Zhang et al., 2019). Ethnic heterogeneity is positively related to paper citations for 2.5 million US papers from 1985 to 2008 (Freeman & Huang, 2015). Similarly, papers with the involvement of professors are more likely to be published in high-quality journals (Bales et al., 2014). As to geographic variables, physical proximity can boost collaborations within the same departments, institutions or countries due to lower transaction costs (Cronin, 2008), but geographically diverse teams generate high impact (Hsiehchen et al., 2015; Wagner et al., 2017). Meanwhile, cross-disciplinary teams have clear division of labor in empirical tasks and can synthesize knowledge from different fields (Haeussler & Sauermann, 2020). In collaboration networks, social ties, structural holes, and prior collaboration experience are vital to team performance. There is an inverted U-shape between social ties and paper citations, which suggests that the combination of weak ties and strong ties can achieve the best performance (Wang, 2016). When assembling teams, newcomers and incumbents jointly determine outcomes in both artistic and scientific teams (Guimera et al., 2005). Transdisciplinary and age-diverse computer science teams have more citations than homogenous teams when they have persistent collaboration (Bu et al., 2018). Finally, team size as the structure of teams has been studied in team collaboration. Larger teams are the major force to develop incremental innovations while small teams are more likely to disrupt science (Wu et al., 2019). The related literature shows that team power dynamics can notably impact team performance but team power has been insubstantially studied in Science of Science. This paper fills the gap by scrutinizing team power dynamics of millions of teams in five disciplines.

## METHODS

*Data*

This research studies team power dynamics in five disciplines: Computer Science (CS), Physics, Sociology, Library & Information Science (LIS), and Art. CS is selected as the focal discipline and is compared with other four disciplines. DBLP is selected to represent CS (Ley, 2009). Papers about Physics, Sociology, LIS, and Art are retrieved from Microsoft Academic Graph (MAG, Wang et al., 2020). Author names have been disambiguated in DBLP (Tang et al., 2012) and MAG (Wang et al., 2020). MAG assigns a field of study for each paper. Field of study is a tree with six layers that its top layer (containing 19 fields) represents the most general fields, such as Physics, Sociology, and Art. The second layer includes 292 fields which are subfields of the first layer. LIS is not a field of study in MAG, but we collect papers with a field of study from the second layer including data science, library science, and information retrieval to represent the LIS field. The order we deal with the CS dataset is we calculate authors' career age in the whole dataset firstly and then we choose the data from 1980 to 2020 because the number of papers before 1980 is too sparse (0.5%, 17,551 papers). Next, we exclude all solo-authored papers and those papers with authors' career age above 80 based on the fact that individual scientists do not have active careers longer than most people's lifespan. We repeat the same steps for all other four disciplines. Table 1 shows the overall statistics of data.

Table 1. Descriptive statistics of five disciplines

|  | Discipline | Year | Co-authored papers | All citations | Unique authors | Mean team power level | Mean Team Power hierarchy | Mean 2-year citations | Mean 5-year citations |
|---|---|---|---|---|---|---|---|---|---|
| DBLP | CS | 1980-2020 | 3,658,127 | 36,286,506 | 3,570,450 | 7.45 | 0.30 | 2.40 | 5.09 |
| MAG | LIS |  | 245,371 | 583,014 | 525,779 | 8.09 | 0.27 | 0.67 | 1.43 |
|  | Physics |  | 3,388,333 | 21,614,368 | 4,317,847 | 9.85 | 0.32 | 1.72 | 3.37 |
|  | Sociology |  | 513,870 | 880,935 | 851,316 | 7.21 | 0.23 | 0.29 | 0.72 |
|  | Art |  | 66,760 | 4,753 | 159,786 | 5.22 | 0.20 | 0.01 | 0.03 |

*Measures*

Career age is the proxy for power in this paper. An author's career age can be measured by the difference between the year of the first publication and the year of the focal publication. For example, a researcher published the first paper in 2015, the focal publication in which he is a team member was published in 2019, his career age is 5 according to the focal publication. In this paper, all coauthors of a paper are treated as a team. There are 3,658,127 CS teams, 245,371 LIS teams, 3,388,333 Physics teams, 513,870 Sociology teams, and 66,760 Art teams. The average career age of a team represents team power level (Greer & van Kleef, 2010). Team power hierarchy is quantified by the Gini coefficient of career ages of all coauthors from a paper (Bunderson et al., 2016; Harrison & Klein, 2007). Table 1 shows that Art has the lowest mean team power level and Physics has the highest mean team power hierarchy. By considering citation inflation and the fact that papers from Social Sciences and Arts & Humanities need longer time to get cited, we select 2-year citation and 5-year citation to measure team impact. CS has the highest 2-year and 5-year citations.

*Factors related to Team Power Dynamics*

The following factors related to team power dynamics are included in this research:

- Team size: Teams are divided into size from 2 to 8+.
- Gender: Gender is assigned based on authors' given name using the name-gender lists from Larivière et al. (2013) which are summarized based on US Census, Wikiname, Wikipedia, and country-specific cases. 84% of given names in CS are matched by applying name lists and 96% of teams have at least one author with an identified gender. Teams are classified into male-dominated, female-dominated, and equal. For example, if identified females outnumber identified males, then the team is female-dominated. In summary, male-dominated teams occupy 75% (2,728,880), female-dominated teams 9% (321,362), and equal teams 12% (450,872) in CS teams.
- Academia vs. Industry: 22,049 organizations from CS papers were assigned to one of the eight categories (i.e., government, education, company, facility, healthcare, nonprofit, archive, and others) defined by the Global Research Identifier Database (GRID) using the methods from Manjunath et al. (2021). Only education (academia) and company (industry) categories were considered here. Since major institutions in healthcare category are universities, medical schools, and research centers, we merge those to the education category. Afterwards, teams are divided into three groups: pure academia, pure industry, and combined. Pure academia dominates CS teams (58%, 2,114,445), followed by pure industry (4%, 135,934) and academia and industry combined teams (2%, 91,221).
- Country: We identify countries based on authors' affiliations. 81% (2,970,336) of CS teams have at least one identified country and 65% (2,368,687) of them come from the same country.

*Multivariable Regression*

Correlation is commonly used to measure the dependency of two variables. However, it can be heavily influenced by confounding factors which can lead to spurious associations. In order to control confounding factors, we use the linear regression model to estimate the relationship between team impact and team power dynamics. Because team impact and team power hierarchy have the curvilinear relationship (see Fig 5), we choose the multivariable regression below (see Equation 1):

$$Team\ Impact_i = \alpha + \beta_1(Team\ Power\ Level_i) + \beta_2(Team\ Power\ Hierarchy_i) + \beta_3(Team\ Power\ Hierarchy_i^2) + \beta_4(Team\ Power\ Level_i \times Team\ Power\ Hierarchy_i) + \beta_5(Team\ Power\ Level_i \times Team\ Power\ Hierarchy_i^2) + \beta_6(Controls) + e_i \quad (1)$$

The dependent variable $Team\ Impact_i$ is the log normal value for 2-year or 5-year citations of a team $i$ as $log(citations + 1)$. The continuous independent variables are $Team\ Power\ Level_i$ and $Team\ Power\ Hierarchy_i$. We calculate the variance inflation factor (VIF) to test the multicollinearity of team power level and team power hierarchy. Their VIF is 1.03, far less than 5 (i.e., the standard judging multicollinearity). Therefore, these two variables are not multicollinear in our model and will not influence the regression results.

$\beta$ represents coefficient for different variables. $\beta_1(Team\ Power\ Level_i)$ represents the linear relationship between team power level and team impact, and $\beta_2(Team\ Power\ Hierarchy_i) + \beta_3(Team\ Power\ Hierarchy_i)^2$ indicates the curvilinear relationship between team power hierarchy and team impact. $\beta_4(Team\ Power\ Level_i \times Team\ Power\ Hierarchy_i) + \beta_5(Team\ Power\ Level_i \times Team\ Power\ Hierarchy_i^2)$ shows the interactive effect of team power level and team power hierarchy with team impact. Controls are variables including team size (*Team Size*), the number of publications of the most productive author in a team (*N of Papers for the Most Productive Author*), gender group of a team (*Male/Female/Equal*), and collaboration type of a team (*Industry/Academia/Combined*). *N of Papers for the Most Productive Author* counts the number of articles published by the most productive author in a team before the year of the focal paper. To control the time-invariant factor, publication year is treated as fixed effect. $e_i$ represents the residual.

## RESULTS AND DISCUSSION

*Descriptive Analysis of Team Power Dynamics*
<u>Gender in CS</u>

For teams in Computer Science, the mean team power levels for male-dominated, female-dominated, and equal teams are 7.67, 6.71, and 7.26, respectively. As to the mean team power hierarchy, male-dominated teams rank first (0.31), followed by female-dominated teams (0.29) and equal teams (0.28). Fig 1 shows the team power level and team power hierarchy of different gender groups in different team sizes and time periods. Team power level of all gender groups decreases while team power hierarchy increases when team gets bigger, and both team power level and team power hierarchy of all gender groups increase over time. Our results reveal that male-dominated teams have more senior researchers and larger career age differences than female-dominated teams. This further signifies that female-dominated teams have comparably younger researchers with smaller career age gaps. The lower degree attainment (Masters and Doctorate) of women in CS[1] coupled with shorter career lengths and higher dropout rates throughout their career (Huang et al., 2020), might lead to a lack of senior female researchers.

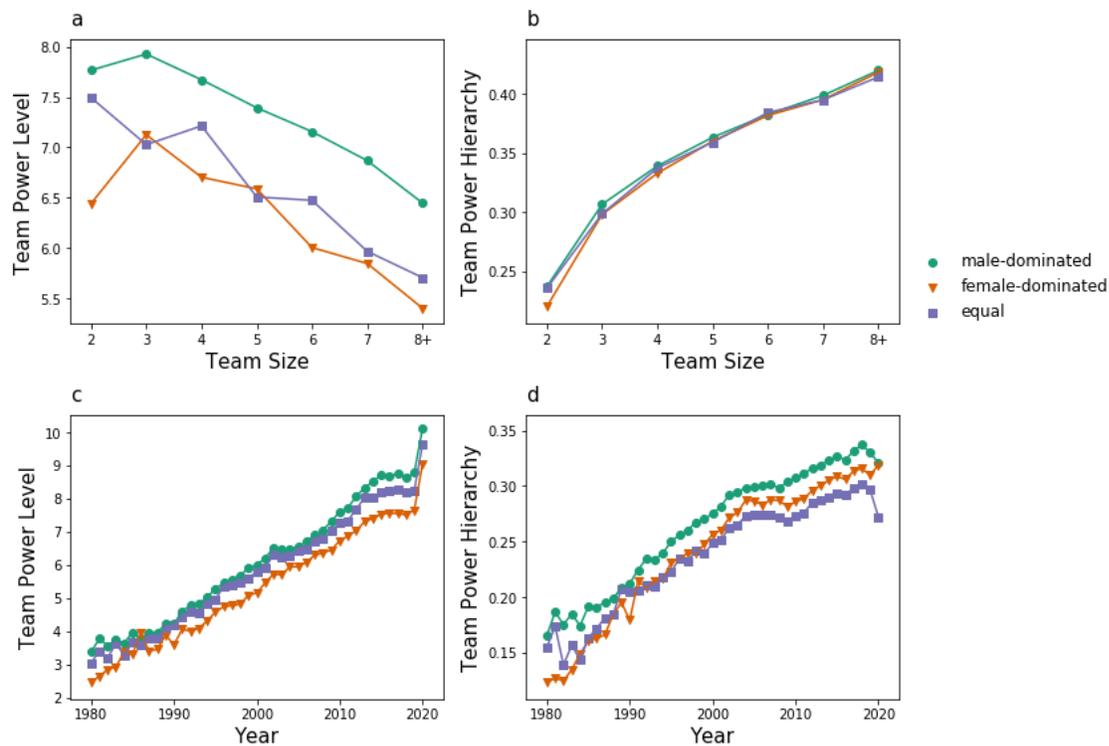

**Fig 1.** Team power level and team power hierarchy across different gender groups. **a.** Team power level of all gender groups descreases when team size increases. **b.** Team power hierarchy in all gender groups increases when team size increases. **c-d.** Team power level and team power hierarchy of all gender groups increase over time. 2020 data are outliers.

Academia vs. Industry in CS

The mean team power level for combined, academia and industry teams in CS are 8.62, 7.75 and 5.79, respectively. Combined teams rank first on mean team power hierarchy (0.34), followed by academia (0.31) and industry teams (0.25). Fig 2 shows all teams' team power level and team power hierarchy increase over time. Industry teams have persistent patterns of lowest team power level and team power hierarchy regardless of team size and time period, while combined teams have the highest team power level and team power hierarchy. In CS, teams from industry and academia collaborations usually have more senior researchers and larger gaps in career age, while teams only from industry contains much younger researchers and smallest career age gaps. When teams get larger, all teams become younger but career age gaps still continue to expand. The difference in team power between pure academia and industry teams in CS could be explained by their distinct publication dynamics (Larivière et al., 2018). Academia teams have higher demands for publications, thus they persist in publishing papers and have more senior researchers.

---
[1] https://www.nsf.gov/statistics/2017/nsf17310/digest/fod-women/computer-sciences.cfm

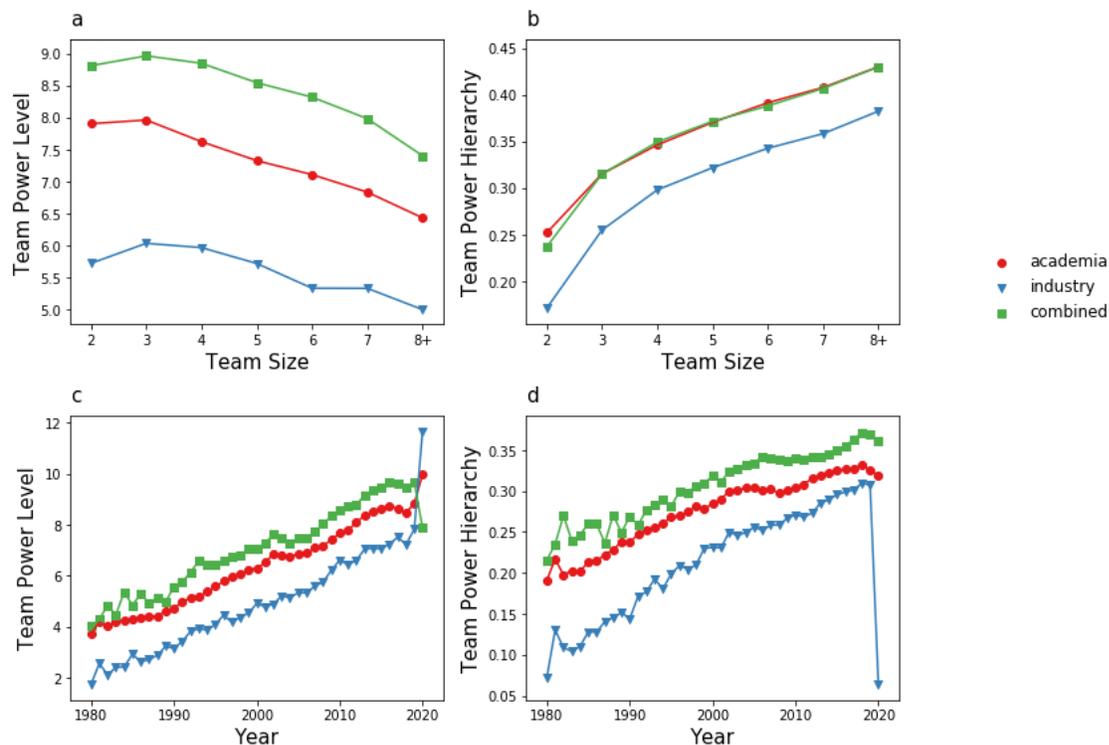

**Fig 2.** Team power level and team power hierarchy across different collaboration groups. **a-b.** Team power level of all groups descreases and team power hierarchy of all groups increases when time size increases. **c-d.** Team power level and team power hierarchy of all groups increase over time. The team power level and team power hierarchy for industry in 2020 are outliners due to limited identified industry teams (data collected until April 2020).

Country in CS

By counting the frequency of all authors' countries, United States (US) and China are the top nations publishing a significant number of CS papers. Thus, we select teams with institutions from China (12%, 445,009) and US (24%, 864,599) for direct comparison. The mean team power levels for US and Chinese teams are 8.40 and 5.41. US teams have higher team power level and team power hierarchy than Chinese teams in different team sizes and time periods (Fig 3). When team size increases, Chinese teams get older while US teams becomes younger (Fig 3a). The difference of team power hierarchy between US (0.32) and China (0.27) exists (Fig 3b) and both increase over time (Fig 3d). It suggests Chinese teams are relatively younger than US teams, having more newcomers and less senior researchers. On one hand, in recent years, China recruits more new CS researchers in the prominent Chinese research institutions, which decrease the team power level (Frank et al., 2019). On the other hand, as a developing country, the brain drain of top senior researchers in China is a severe problem (Yuan et al., 2020).

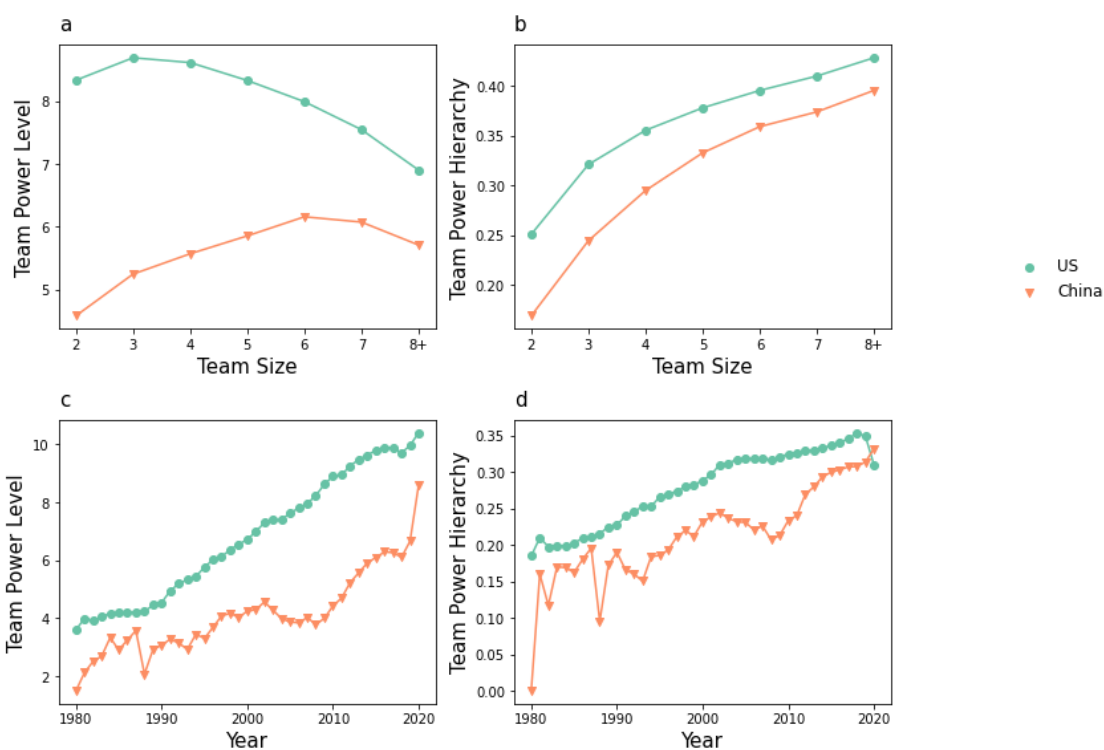

**Fig 3.** Team power level and team power hierarchy across US and Chinese teams. **a-b.** Team power level and team power hiearchy of US and Chinese teams over team size. **c-d.** Team power level and team power hierarchy of both groups increase over time.

Disciplines

We extend the team power analysis from CS to Physics, Sociology, Art, and LIS. Changes of team power level and team power hierarchy across different team sizes and time periods among these five disciplines are shown in Fig 4. When team size gets larger (Fig 4a), CS teams' team power level decreases while teams from Physics, LIS, and Art increase and Sociology stay flat, but all five disciplines show the increase trend of team power hierarchy (Fig 4b). This indicates that large teams contain more senior researchers and the career age gaps expand, except CS teams. This is because large teams in CS consist of more researchers with low career ages and the differences of their career ages have been enlarged (Fig S1). It is unique for CS teams since large teams require more young scholars to do technical tasks. Team power level and team power hierarchy of CS, LIS, and Physics show the steady increase trend, while Sociology has the slow increase but Art stays no change or even has slight decrease over time. As to CS, the steady increase is because when new researchers are increasing (career age = 1), senior researchers are also increasing (career age > 2) (Fig S2). These patterns in different disciplines show that Science teams tend to have hierarchical structure compared with Sociology or Art teams. The practice of collaboration is different in different disciplines. The students in science rely more on their supervisors in research directions and they need to work with senior researchers. By contrast, researchers in social science relatively have more independence in their research and they can work with peers (Moody, 2004).

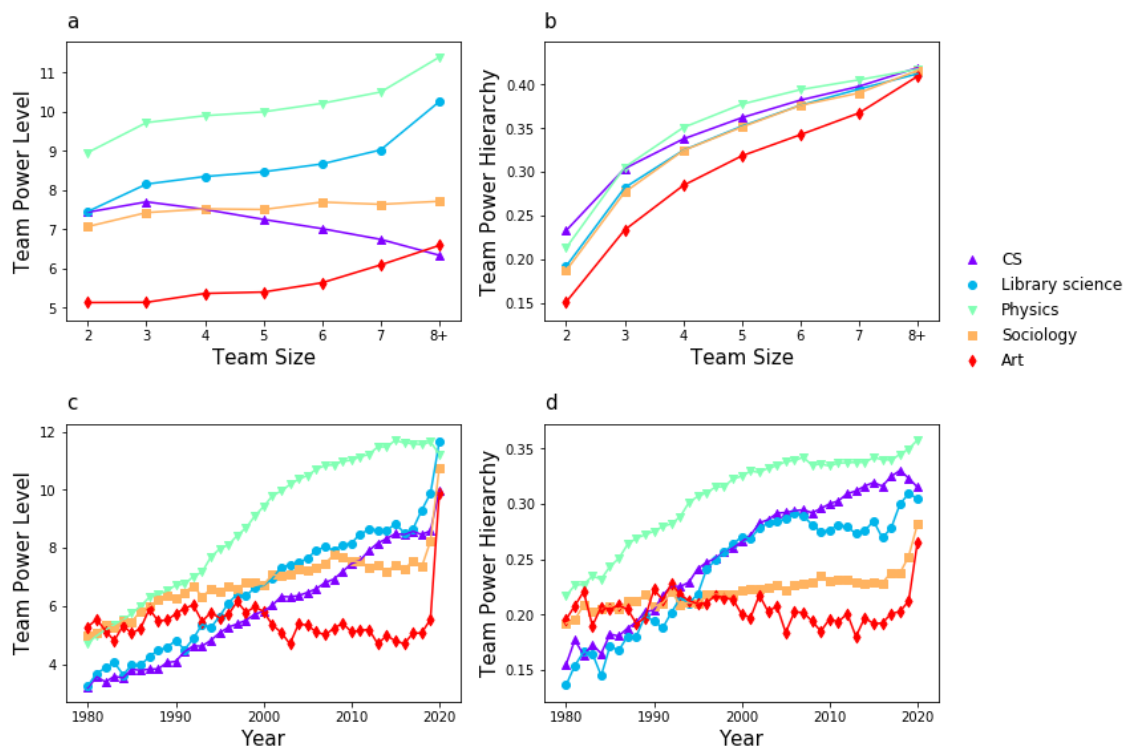

**Fig 4.** Team power level and team power hierarchy across five disciplines. Team power level and team power hierarchy with team size (**a-b**) and time (**c-d**).

*Team Power Dynamics and Team Impact in CS*

Fig 5a shows that teams with high team power level are cited more on 2-year citations. While team power hierarchy and team impact demonstrate a curvilinear relationship (Fig 5b), it indicates that high team power hierarchy can hurt team impact severely. Figure 5c-d reveals the combination of team power level and team power hierarchy on team impact measured by 2-year and 5-year citations. We rank teams' team power level in an ascending order and calculate the percentage of team power level that 20% means those teams with the lower 20% of team power level, same for team power hierarchy percentage. Team power level and team power hierarchy are divided into 1-10 based on the percentiles. We divide all team power hierarchies into two categories based on the turning point of the inverted U-shape between team power hierarchy and team impact (see Fig 5b): teams with high Gini (i.e., Gini coefficients > turning point) and teams with low Gini (i.e., Gini coefficient < turning point). Because of the curvilinear relationship between team power hierarchy and team impact, using turning point to divide groups into low Gini and high Gini can guarantee the linearity of both groups. Dots in Fig 5c-d represent the citations of teams using the gradient of green color to indicate high (dark) or low (light) team power hierarchy. K is the slope of fitting lines of teams with high Gini and low Gini, and the slopes of both teams are significantly different through Kolmogorov-Smirnov (KS) test ($p < 0.001$). Teams with low team power hierarchy (Low Gini solid line) have more citations than teams with high team power hierarchy (High Gini dotted line). When team power level increases, the difference in citations between low Gini teams and high Gini teams also increases. It confirms that flat team structure increases team impact, especially when team members have high career ages on average. In summary, a clear pattern has been identified in CS that: 1) Team power level has a positive correlation with team impact; 2) Team power hierarchy has a curvilinear relationship with team impact: and 3) The combination of team power level and team power hierarchy does affect team impact that when team power level increases, teams with low team power hierarchy have higher team impact than teams with high team power hierarchy. We found that when we extend the citation window after 10 years, our finding is consistent (Fig S3). The gap in after 10-year citations is more pronounced between flat structure (Low Gini) and hierarchical structure (High Gini).

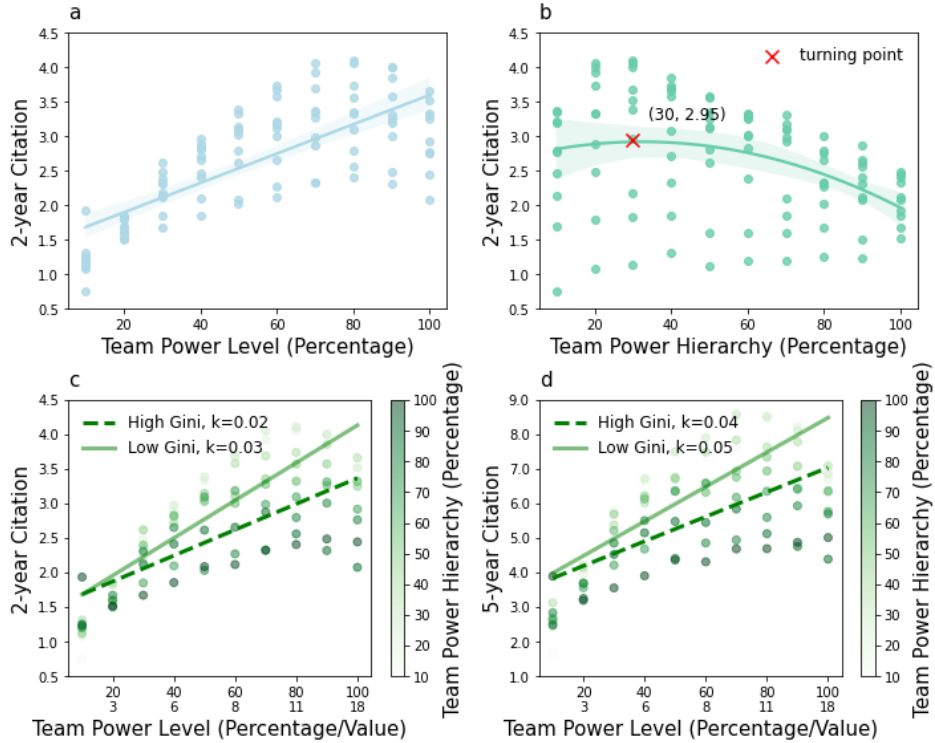

**Fig 5.** Team power level and team power hierarchy on team impact in CS teams. **a-b.** A fitting line was drawn based on the scatter plot of team power level and 2-year citations. Team power hierarchy has a curvilinear relationship with 2-year citations. **c-d.** The combination effect between team power level and team power hierarchy on team impact measured by 2-year and 5-year citations. The turning point for team power hierarchy is 30% for 2-year citations and 20% for 5-year citations. A gradient green color indicates that darker sides suggest higher team power hierarchy. Both percentile and corresponding value of team power level are shown in **c-d**.

We use the multivariable regression model (see Equation 1) to include control variables which contain team size, the number of countries of a team, the number of papers for the most productive author in a team, industry/academia/combined team, and male/female/equal team, and time as fixed effect. The coefficients of the interactive effect between team power dynamics (e.g., team power level and team power hierarchy) and team impact are significant ($p < 0.001$). The coefficients for $Team\ Power\ Level_i \times Team\ Power\ Hierarchy_i$ are negative and those for $Team\ Power\ Level_i \times Team\ Power\ Hierarchy_i^2$ are positive (see Table 2). In Fig 6, we divide teams into three categories based on team power level: teams with the mean of all team power levels (called Mean Teams), teams with the mean+std (called Mean+Std Teams), and teams with the mean-std (called Mean-Std Teams). All three team categories reach to their turning points on 2-year and 5-year citations when team power hierarchy is low. When team power hierarchy increases after the turning points, team impact drops dramatically for all three categories. Meanwhile, Mean+Std teams reach the turning points with the lowest team power hierarchy compared with Mean teams and Mean-Std teams. The interactive effect between team power level and team power hierarchy on team impact visualized in Fig 6 confirms that the flat structure is associated with higher team impact. In Table 2, the number of countries of a team has a positive relationship with citations. The publications of the most productive author of a team bring more citations for teams. Compared with academia teams, industry teams are significantly associated with an average increase of 13% in log citations within 2 years and 16% in log citations within 5 years, while combined teams have an average increase of 18% in log citations within 2 years and 23% in log citations within 5 years. Compared with male teams, female-dominated teams receive 4% less and equal teams receive 3% less in log citations within 2 years and 5 years. The same pattern has been identified: 1) team power level is positively related to team impact; 2) team power hierarchy has a curvilinear relationship with team impact; and 3) When team power level increases, teams with low team power hierarchy have higher team impact than teams with low team power hierarchy, which verifies that teams with flat structure have high team impact. These results are consistent in two main countries' teams, China and US (Fig S4). In this paper, we use career age as the proxy for power. But there are other ways to measure power. For example, the number of publications of an author indicate the expertise/knowledge that this author possessed; while the number of collaborators show the social capital that this author accumulated. We use the mean and Gini coefficient of the number of collaborators of each author in a team to represent its team power level and team power hierarchy, same for the number of publications. Using different measures of power, we still get the same results that teams with flat structure achieve high impact (Table S1).

Table 2: Multivariable Regression for Team Power Dynamics and Team Impact in CS

|  | CS | |
| --- | --- | --- |
|  | 2-Year | 5-Year |
| Team Power Level (Mean) | 0.02*** (0.00) | 0.03*** (0.00) |
| Team Power Hierarchy (Gini) | 1.79*** (0.14) | 2.42*** (0.02) |
| Team Power Hierarchy $^2$ | -2.96*** (0.03) | -3.85*** (0.04) |
| Team Power Level × Team Power Hierarchy | -0.05*** (0.00) | -0.08*** (0.00) |
| Team Power level × Team Power Hierarchy $^2$ | 0.05*** (0.00) | 0.08*** (0.00) |
| Team Size | 0.03*** (0.00) | 0.03*** (0.00) |
| N of Countries | 0.06*** (0.00) | 0.08*** (0.00) |
| N of Papers for the Most Productive Author | 0.001*** (0.00) | 0.002*** (0.00) |
| Academia/Industry/Combined | | |
| Industry | 0.13*** (0.00) | 0.16*** (0.00) |
| Combined | 0.18*** (0.00) | 0.23*** (0.00) |
| Male/Female/Equal | | |
| Female | -0.04*** (0.00) | -0.04*** (0.00) |
| Equal | -0.03*** (0.00) | -0.03*** (0.00) |
| N of Countries | 0.06*** (0.00) | 0.08*** (0.00) |
| N of Papers for the Most Productive Author | 0.001*** (0.00) | 0.002*** (0.00) |
| Year fixed effect | Yes | |
| $R^2$ | 0.12 | 0.16 |
| N | 3,658,127 | |

Note. * $P < 0.05$; ** $P < 0.01$; *** $P < 0.001$; 2-year means 2-year citation, 5-year means 5-year citation

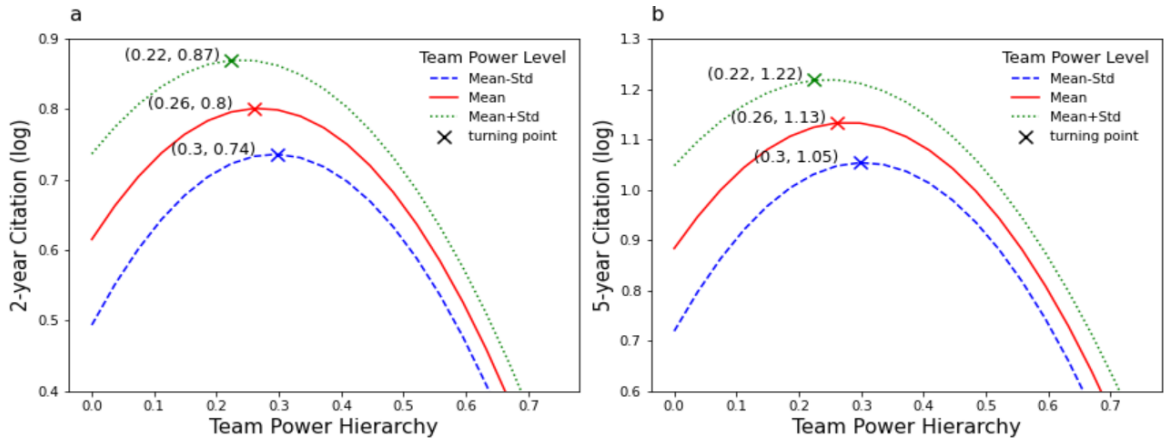

**Fig 6**. The visualization between team power dynamics and team impact in CS after controlling confounding variables in multivariable regression. We draw dashed dotted green line (Mean+Std teams), solid red line (Mean teams) and dashed green line (Mean-Std teams) in **a-b** to show the relationship between team power hierarchy and 2-year or 5-year citation at different team power levels.

*Team Power Dynamics and Team Impact in Other Disciplines*

The multivariable regression (Equation 1) with team size as control variable and time as fixed effect has been used to compare the interactive effect of team power dynamics on team impact for LIS, Physics, Sociology, and Art (Table 3). Fig 7 presents the combination of team power level and team power hierarchy on team impact. Similar patterns identified in CS have been repeated in Physics (Fig 7c-d) and LIS (Fig 7a-b), are less significant in Sociology (Fig 7e-f), and do not exist in Art (Fig 7g-h) that when team power level increases, teams with low team power hierarchy get more citations. Flat team structure is associated with higher team impact except in Art.

Table 3. Multivariable regression analysis of team power dynamics and team impact in different disciplines

|  | LIS (1) | | Physics (2) | | Sociology (3) | | Art (4) | |
| --- | --- | --- | --- | --- | --- | --- | --- | --- |
|  | 2-Year | 5-Year | 2-Year | 5-Year | 2-Year | 5-Year | 2-Year | 5-Year |
| Team Power Level (Mean) | 0.004*** (0.00) | 0.005*** (0.00) | 0.02*** (0.00) | 0.02*** (0.00) | 0.01*** (0.00) | 0.01*** (0.00) | 0.00*** (0.00) | 0.00*** (0.00) |
| Team Power Hierarchy (Gini) | 1.09*** (0.03) | 1.61*** (0.04) | 1.21*** (0.01) | 1.66*** (0.01) | 0.40*** (0.01) | 0.75*** (0.02) | 0.01*** (0.00) | 0.02*** (0.00) |
| Team Power Hierarchy $^2$ | -1.61*** (0.06) | -2.37*** (0.08) | -1.71*** (0.02) | -2.29*** (0.03) | -0.56*** (0.03) | -1.02*** (0.05) |  |  |
| Team Power Level × Team Power Hierarchy | 0.01** (0.00) | 0.01** (0.00) | -0.01*** (0.000) | -0.02*** (0.00) | 0.004** (0.00) | 0.004* (0.00) | -0.00 (0.00) | -0.00 (0.00) |
| Team Power level × Team Power Hierarchy $^2$ | -0.02** (0.01) | -0.02* (0.01) | 0.02*** (0.00) | 0.03*** (0.00) | -0.02*** (0.00) | -0.03*** (0.00) |  |  |
| Team Size | 0.00 (0.00) | 0.00 (0.00) | 0.00*** (0.00) | 0.00*** (0.00) | 0.001* (0.000) | -0.001 (0.00) | -0.00 (0.000) | -0.00* (0.00) |
| Year fixed effect | Yes | Yes | Yes | Yes | Yes | Yes | Yes | Yes |
| $R^2$ | 0.11 | 0.16 | 0.12 | 0.16 | 0.09 | 0.15 | 0.004 | 0.01 |
| N | 245,371 | | 3,388,333 | | 513,870 | | 66,760 | |

Note. * $P < 0.05$; ** $P < 0.01$; *** $P < 0.001$; 2-year means 2-year citation, 5-year means 5-year citation

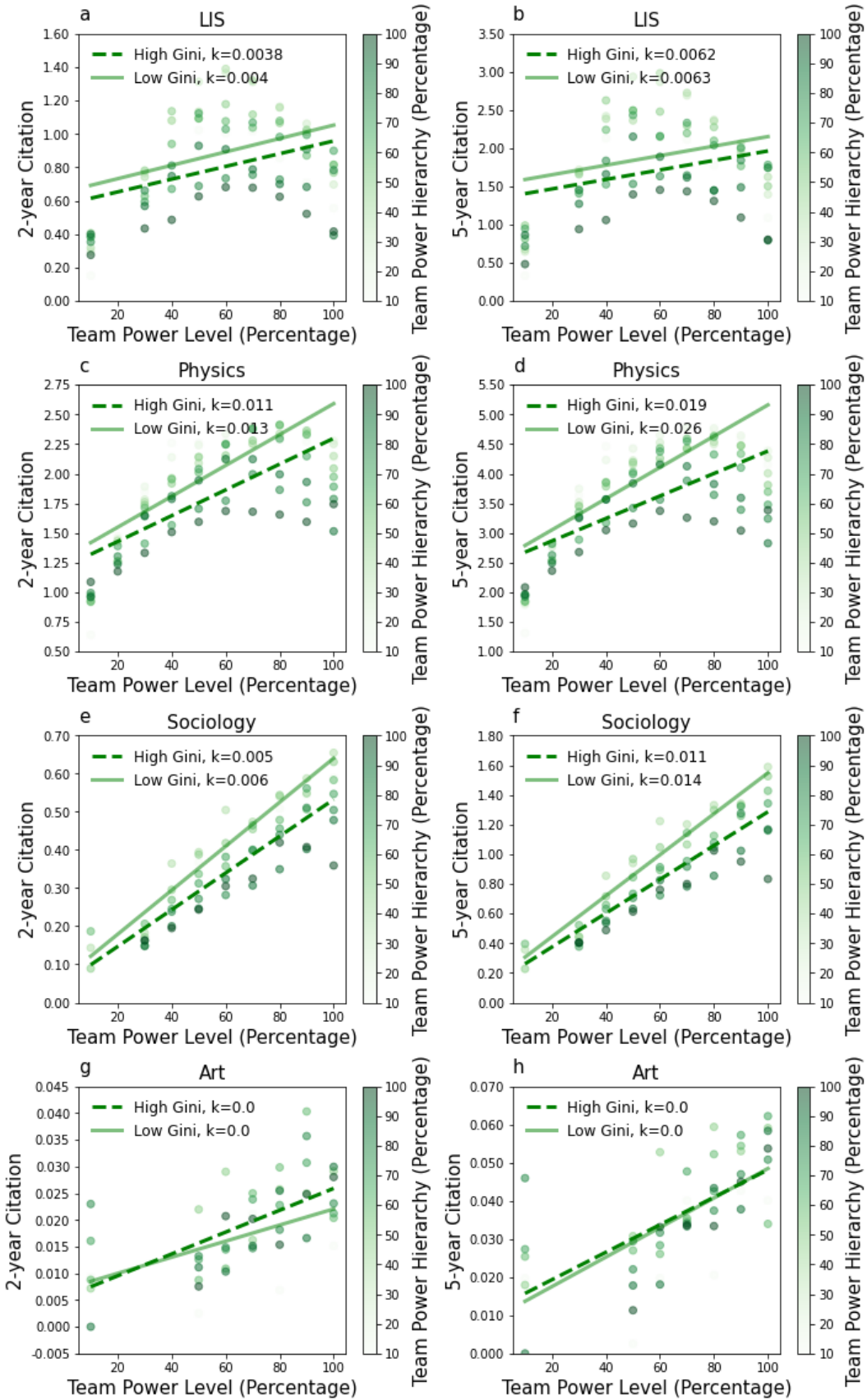

**Fig 7.** The effect of team power level and team power hierarchy on team impact from LIS, Physics, Sociology and Art. Teams are divided into high Gini and low Gini based on the turning point of LIS (30%), Physics (20%), and Sociology (40%) within both 2-year and 5-year citations. Art has no turning point, therefore median of all Gini coefficients is used to form high Gini and low Gini groups (**g-h**).

## CONCLUSION

Power in scientific teams can be manifested through seniority (e.g., career age), productivity (e.g., the number of publications), and collaboration (e.g., the number of coauthors). This paper quantifies team power level and team power hierarchy using authors' career age to understand the characteristics of team power dynamics and their interactive effects on team impact. It analyzes various factors related to team power including team size, time, collaboration type (e.g., academia/industry/combined), gender (e.g., male/female/equal), and country (e.g., US vs. China). In CS, team power hierarchy decreases but team power level increases when team size increases; team power level and team power hierarchy increase over time; male-dominated teams have higher team power level, higher team power hierarchy, and receive more citations than female-dominated and equal teams; combined teams have higher team power level, higher team power hierarchy, and receive more citations than academia and industry teams; and US teams have higher team power level and higher team power hierarchy than Chinese teams. When comparing CS with Physics, LIS, Sociology, and Art, some general patterns have been identified that: 1) Team power level has a positive correlation with team impact for all five disciplines; 2) Team power hierarchy has a curvilinear correlation with team impact for all fields except Art; and 3) When team power level increases, teams with low team power hierarchy get higher team impact than teams with high team power hierarchy, except Art. Flat structure is associated with better team impact. Our findings are consistent with the outcomes of Greer and Van Kleef (2010) and Greer et al. (2011) that their contexts are similar as the context of scientific activities which are knowledge intensive, demanding flat structure to create open space for innovation, but are different from findings from Bloom (1999) and Goodwin et al. (2018) where their contexts are military and sports.

Scientific research contains knowledge intensive activities which require teams to efficiently share knowledge. Mazorodze and Buckley (2019) found that a flat structure in knowledge-intensive organizations supports knowledge sharing by efficiently leveraging existing expertise, experiences, and competencies though borderless communication. Flat team structure can eliminate organizational layers and create decentralized environments to enable effective knowledge sharing that existing knowledge and created new knowledge can flow freely among individual team members (Riege, 2007). Team members with similar career ages are more likely to have shared mindset and joint responsibility which can remove the barrier of communication (Haas & Mortensen, 2016). As Haas and Mortensen pointed out that the success of collaboration still depends on the fundamentals that what matters most is not the personalities or behavior of team members but whether a team has a strong structure and a supportive context, flat structure and boundless communication can be easily established for team members with similar ages because people with different ages have distinct communication preferences (Clarke et al., 2020).

The benefits of flat team structure which are supported by rich literature include: 1) adapting to unpredicted and creative tasks, 2) creating a psychologically satisfaction; and 3) supporting coordination. Flat structure is crucial to success when tasks are changing, complex, ambiguous and require much creativity, whereas hierarchical structure can benefit teams when tasks are relatively stable, simple, predictable and require little creativity (Anderson & Brown, 2010). Faced with such kind of tasks, egalitarian teams enable more people to participate in making decisions (Duncan, 1973), aggregate the wisdom across individuals (Surowiecki, 2002), and increase the heterogeneity of opinions (Gruenfeld & Tiedens, 2010). Scientific research is non-routine and complicated, and it requires demonstrating novelty by advancing the state of art. In the scientific context, Xu et al. (2022) found that flat teams are more novel in ideas and disrupt prior research than hierarchical teams. In psychology, flat structure can boost the satisfaction of group members, thus reducing the possibility of turnover (Griffeth et al., 2000). Flat structure can increase levels of motivation for the whole group rather than the minority. In the hierarchical structure, the inequality might make low-ranking members feel unfair but they lack the ability to change their status and thus become less motivated (Magee & Galinsky, 2008). It is crucial to retain and motivate new researchers in science. Flat structure can boost communication and cooperation. Equal turn-taking communication in the flat structure can achieve better team performance (Woolley et al., 2010). This study contributes to the science of teamwork by providing quantitative evidences to support flat structure for effective team collaboration through analyzing the relationship between team power dynamics and team impact using 7.7 million teams from five disciplines. It suggests practical implications that setting up flat team structure is important for better team performance.

There are several limitations in this research. First, career age is a simple proxy of power in scientific teams. We must admit there are few exceptions that young researchers can hold senior positions and show their superiority in creativity. Since power can be represented with position, expertise, ability, and perceived intelligence (Greer & Van Kleef, 2010; Groysberg et al., 2011; Piercy, 2019), we can consider using other alternative measures, like h-index to measure team power. Second, we use the citation number to evaluate team impact which has limitations due to citation inflation and inherited biases. Citation counts are not adequate indicators of research value (Sugimoto et al., 2021). It would be interesting to measure success using disruptive innovation (e.g., new knowledge overshadowing past ideas, Wu et al. 2019) or combinational novelty (e.g., atypical combination of pre-existing knowledge components, Uzzi et al., 2013). Thirdly, author name disambiguation methods might have problems to identify early researchers due to poor metadata kept in earlier years. Meanwhile, since the author name disambiguation has limitations in identifying east-Asian names, it is possible that an experienced scholar might be identified as multiple new scholars, which might lead to low team power level and team power hierarchy in Chinese teams compared with US teams. Finally, as to the DBLP dataset, papers before 1980 are relatively spare. A possible consequence is that the DBLP dataset lacks complete bibliometrics history of authors' prior publications before 1980. Although 1980 is a key time point which marks the exponential increase of CS papers (Fig S5a) and stable team power level and hierarchy changes (Fig S5b-c), we need to acknowledge we do not analyze the whole dataset and the time point cut might cause potential bias. This paper proposes quantitative measures of team power which open the doors for future studies. For example, we can dive deep to figure out why flat structure is better than hierarchical structure, revealing the mechanisms behind the fact. Combining quantitative analyses of team power with qualitative surveys and observations to understand the behavioral and psychological factors is the interesting next step. Also, we can use team power dynamics to understand shared leadership of a team to see whether gender or culture dominated teams will have different team power dynamics. Furthermore, we can revisit academic mentorship from the angle of team power to identify whether the team power in mentors' collaboration network (e.g., the number of collaborators, the number of productive co-authors, and the number of highly cited co-authors) will affect the early stage careers of the mentees.

**ACKNOWLEGEMENT**

Gruenfeld, D. H., & Tiedens, L. Z. (2010). Organizational preferences and their consequences. In S. T. Fiske, D. Gilbert, & G. Lindzey (Eds.), Handbook of social psychology (5th edition, pp. 1252–1286). Hoboken, NJ: Wiley & Sons.

Guimera, R., Uzzi, B., Spiro, J., & Amaral, L. A. N. (2005). Team assembly mechanisms determine collaboration network structure and team performance. Science, 308(5722), 697-702.

Haas, M., & Mortensen, M. (2016). The secrets of great teamwork. Harvard Business Review, June, https://hbr.org/2016/06/the-secrets-of-great-teamwork

Haeussler, C., & Sauermann, H. (2020). Division of labor in collaborative knowledge production: The role of team size and interdisciplinarity. *Research Policy*, *49*(6), 103987.

Halevy, N., Chou, E. Y., Galinsky, A. D., & Murnighan, J. K. (2012). When hierarchy wins: Evidence from the national basketball association. Social Psychological and Personality Science, 3(4), 398-406.

Harrison, D. A., & Klein, K. J. (2007). What's the difference? Diversity constructs as separation, variety, or disparity in organizations. Academy of management review, 32(4), 1199-1228.

He, B., Ding, Y., Tang, J., Reguramalingam, V., & Bollen, J. (2013). Mining diversity subgraph in multidisciplinary scientific collaboration networks: A meso perspective. *Journal of Informetrics*, 7(1), 117-128.

Hsiehchen, D., Espinoza, M., & Hsieh, A. (2015). Multinational teams and diseconomies of scale in collaborative research. Science advances, 1(8), e1500211.

Huang, J., Gates, A. J., Sinatra, R., & Barabási, A. L. (2020). Historical comparison of gender inequality in scientific careers across countries and disciplines. *Proceedings of the National Academy of Sciences*, *117*(9), 4609-4616.

Larivière, V., Desrochers, N., Macaluso, B., Mongeon, P., Paul-Hus, A., & Sugimoto, C. R. (2016). Contributorship and division of labor in knowledge production. Social Studies of Science, 46(3), 417-435.

Larivière, V., Gingras, Y., & Sugimoto, C.R. (2014). Team size matters: Collaboration and scientific impact since 1900. Journal of the Association for Information Science and Technology, 66(7), 1323-1332.

Larivière, V., Macaluso, B., Mongeon, P., Siler, K., & Sugimoto, C. R. (2018). Vanishing industries and the rising monopoly of universities in published research. PloS one, 13(8), e0202120.

Larivière, V., Ni, C., Gingras, Y., Cronin, B., & Sugimoto, C. R. (2013). Bibliometrics: Global gender disparities in science. Nature News, 504(7479), 211.

Lebeau, L. M., Laframboise, M. C., Larivière, V., & Gingras, Y. (2008). The effect of university–industry collaboration on the scientific impact of publications: the Canadian case, 1980–2005. *Research Evaluation*, *17*(3), 227-232.

Ley, M. (2009). DBLP: Some Lessons Learned. Proceedings of the VLDB Endowment. 2 (2). pp. 1493–1500

Linton, R. (1936). The study of man. New York: Appelton-Century.

Magee, J. C., & Galinsky, A. D. (2008). Social hierarchy: The self-reinforcing nature of power and status. Academy of Management annals, 2(1), 351-398

Manjunath, A., Li, H., Song, S., Zhang, Z., & Kumar, I. (2021). Comprehensive analysis of 2.4 million patent-to-research citations maps the biomedical innovation and translation landscape. Nature Biotechnology, 39(6), 678-683.

## Supplementary Materials

Team Size and Time Dynamics in CS

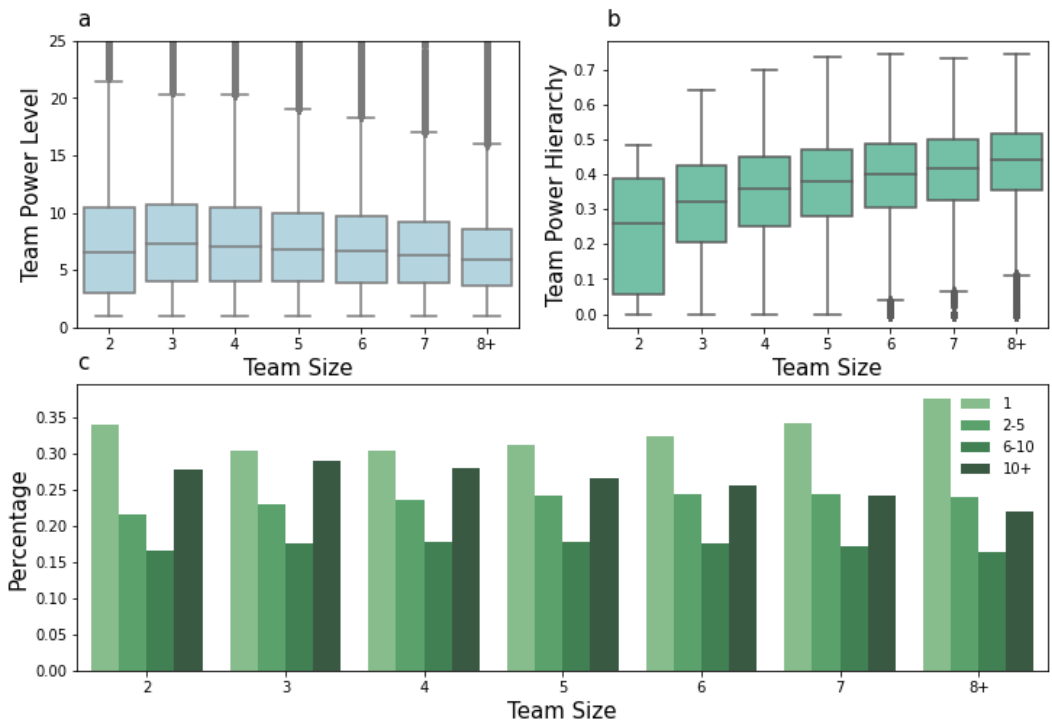

**Fig S1.** Team power dynamics and team size. **a-b.** Box plot of team power level and team power hierarchy for teams in different team size. When teams have more than three researchers, team power level decreases but team power hierarchy increases. **c.** The percentage of researchers in different age groups (career age =1, 2-5, 6-10, >10) in different team sizes.

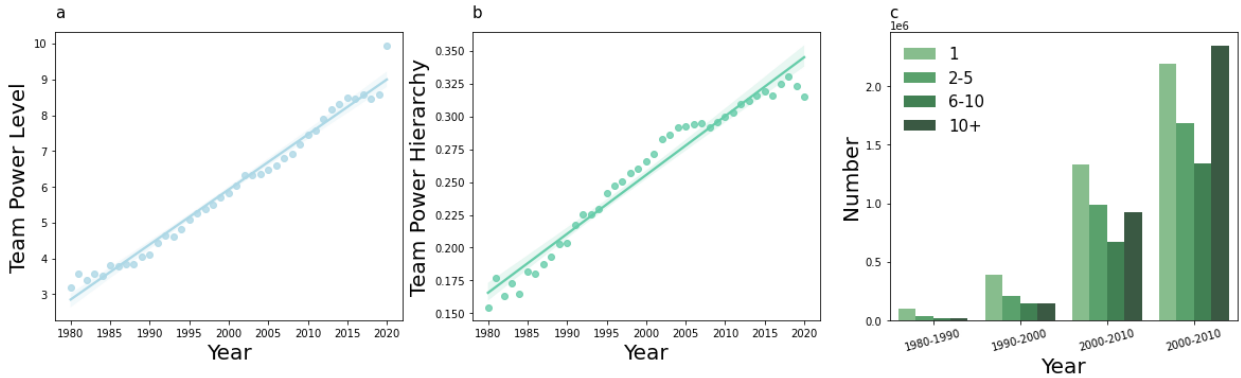

**Fig S2.** Team power dynamics and time. **a.** Mean team power level increases over time. **b.** Mean team power hierarchy increases over time. **c.** The number of researchers in different age groups (career age =1, 2-5, 6-10, >10) in four time periods.

Long Team Impact in CS

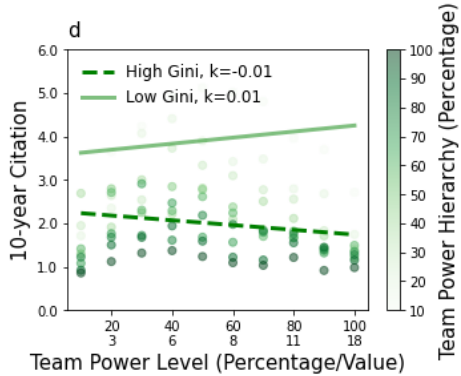

**Fig S3.** Team power level and team power hierarchy on team impact in CS teams. The combination effect between team power level and team power hierarchy on team impact measured by after 10-year citations. The turning point for team power hierarchy is 10% for after 10-year citations. A gradient green color indicates that darker sides suggest higher team power hierarchy. Both percentile and corresponding value of team power level are shown in X-axis.

US vs. China

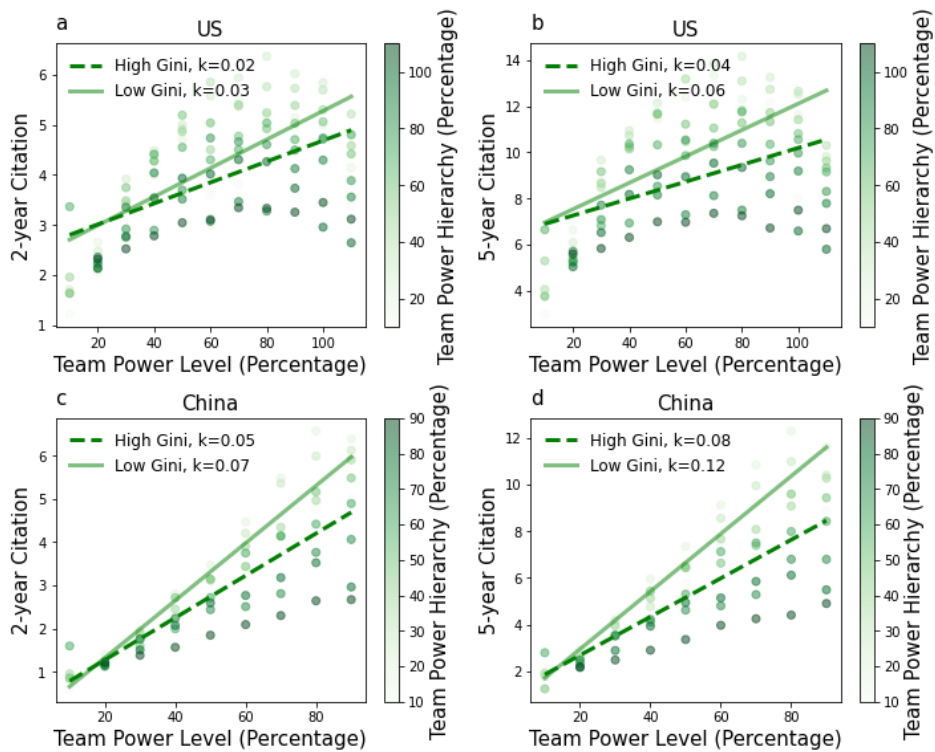

**Fig S4.** The combination of team power level and team power hierarchy on team impact for teams from US and China. **a-b.** US teams on 2-year and 5-year citations. **c-d.** Chinese teams on 2-year and 5-year citations. Teams are divided into high Gini and low Gini based on the turning point which is 30% for US teams (**a-b**) and 20% for Chinese teams (**c-d**) for both 2-year and 5-year citations.

Team Power in CS: Career Age, Publication, and Collaborator

Table S1. Three power measures in CS teams

|  | Career Age | | #Publications | | #Collaborators | |
|---|---|---|---|---|---|---|
|  | 2-Year | 5-Year | 2-Year | 5-Year | 2-Year | 5-Year |
| Team Power level (Mean) | 0.03*** | 0.04*** | 0.03*** | 0.04*** | 0.03*** | 0.04*** |
|  | (0.00) | (0.00) | (0.00) | (0.00) | (0.00) | (0.00) |
| Team Power Hierarchy (Gini) | 1.89*** | 2.87*** | 1.19*** | 1.82*** | 1.25*** | 1.95*** |
|  | (0.01) | (0.02) | (0.01) | (0.02) | (0.01) | (0.02) |
| Team Power Hierarchy $^2$ | -2.89*** | -4.55*** | -1.00*** | -1.67*** | -1.06*** | -1.91*** |
|  | (0.03) | (0.04) | (0.02) | (0.03) | (0.02) | (0.03) |
| Team Power level × Team Power Hierarchy | -0.02*** | -0.09*** | -0.04*** | -0.06*** | -0.03*** | -0.06*** |
|  | (0.00) | (0.00) | (0.00) | (0.00) | (0.00) | (0.00) |
| Team Power level × Team Power Hierarchy $^2$ | 0.01 | 0.12*** | 0.03*** | 0.03*** | -0.00 | 0.01 |
|  | (0.00) | (0.01) | (0.00) | (0.00) | (0.003) | (0.00) |
| $R^2$ | 0.04 | 0.04 | 0.04 | 0.04 | 0.04 | 0.04 |
| N | 3,658,127 | | | | | |

Note. * $P < 0.05$; ** $P < 0.01$; *** $P < 0.001$; 2-year means 2-year citation, 5-year means 5-year citation

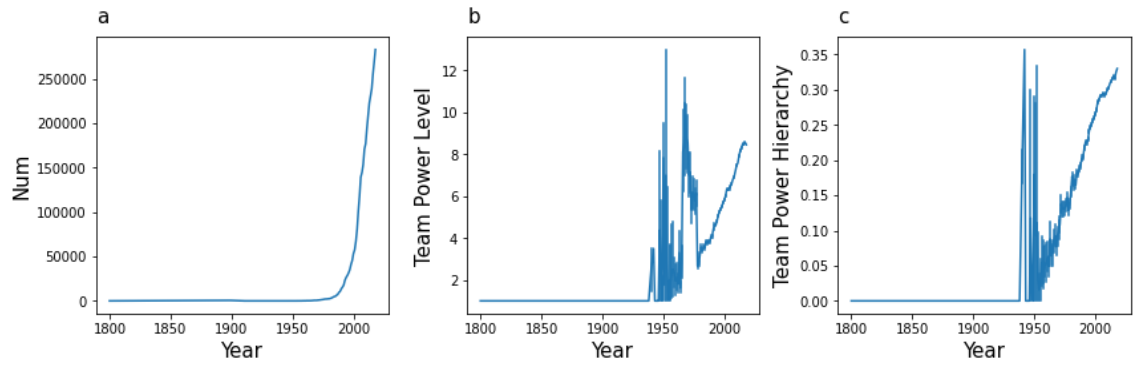

**Fig S5**. Team dynamics from 1800 to 2020 in the whole datatset **a.** The number of teams over time (2019 and 2020 data are incomplete so we do not draw the two time points) **b.** Teams power level dynamcis **c.** Team power hierarchy dynamics.